\documentclass[aps,prd,twocolumn,showpacs,superscriptaddress,superscriptaddress,10pt,longbibliography]{revtex4-1}
\usepackage{graphicx,dcolumn,bm,braket,soul,color,amssymb,MnSymbol,hyperref}
\usepackage{natbib}
\pdfoutput=1

\begin{document}

\title{Ultimate precision: Gaussian parameter estimation in flat and curved spacetime}

\author{Dominik \v Safr\'anek}
\email{dominik.safranek@univie.ac.at}
\affiliation{School of Mathematical Sciences, University of Nottingham, University Park,
Nottingham NG7 2RD, United Kingdom}
\affiliation{Faculty of Physics, University of Vienna, Boltzmanngasse 5, 1090 Vienna, Austria}
\author{Jan Kohlrus}
\affiliation{School of Mathematical Sciences, University of Nottingham, University Park,
Nottingham NG7 2RD, United Kingdom}
\author{David Edward Bruschi}
\affiliation{Racah Institute of Physics and Quantum Information Science Centre, the Hebrew University of Jerusalem, Givat Ram, 91904 Jerusalem, Israel}
\affiliation{York Centre for Quantum Technologies, Department of Physics, University of York, Heslington, YO10 5DD York, United Kingdom}
\author{Antony R. Lee}
\affiliation{School of Mathematical Sciences, University of Nottingham, University Park,
Nottingham NG7 2RD, United Kingdom}
\author{Ivette Fuentes}\thanks{Previously known as Fuentes-Guridi and Fuentes-Schuller.}
\affiliation{School of Mathematical Sciences, University of Nottingham, University Park,
Nottingham NG7 2RD, United Kingdom}
\affiliation{Faculty of Physics, University of Vienna, Boltzmanngasse 5, 1090 Vienna, Austria}

\begin{abstract}
Relativistic quantum metrology provides an optimal strategy for the estimation of parameters encoded in quantum fields in flat and curved spacetime. These parameters usually correspond to physical quantities of interest such as proper times, accelerations, gravitational field strengths, among other spacetime parameters. The precise estimation of these parameters can lead to novel applications in gravimeters, spacetime probes and gravitational wave detectors. Previous work in this direction only considered pure probe states. In realistic situations, however, probe states are mixed. In this paper, we provide a framework for the computation of optimal precision bounds for mixed single- and two-mode Gaussian states within quantum field theory. This enables the estimation of spacetime parameters in case the field states are initially at finite temperature. \end{abstract}

\maketitle
\section{Introduction}
The main aim of quantum metrology is to provide optimal strategies to estimate physical quantities of interest exploiting quantum properties. Usually, these quantities do not correspond to observables of the system, but rather to real parameters, such as time or field strengths, that are encoded on initial probes states through the evolution of the system. The scheme requires that infinitesimally close quantum states can be distinguished after the parameter has been encoded. Therefore, it is necessary to determine what are the optimal measurements that enable to distinguish the states more accurately.  The quantum Fisher information is a function that quantifies how well states can be distinguished by optimal measurements. The ultimate limit of precision for estimating the parameter is given by the so-called quantum Cram\'{e}r-Rao bound, which depends on the quantum Fisher information and the number of measurements made on $N$ identical copies of the state~\cite{Holevo2011a,Petz2010a,Giovannetti2006a}.

Recently, it has become of great interest to exploit quantum metrology techniques to measure gravitational parameters. Novel applications could not only provide important improvements in seismology and oil exploration, but also provide insights on fundamental questions in the overlap of quantum theory and relativity. Our understanding of quantum phenomena in the presence of spacetime remains rather limited. This is not surprising, incompatibilities arise since quantum theory assumes time to be absolute, while in general relativity time is an observer dependent quantity.  Quantum field theory in curved spacetime is a theory that succeeds at describing phenomena in the overlap of quantum theory and relativity by considering that in some regimes spacetime is a classical background underlying the dynamics of quantum fields~\cite{Benini2013a,Hollands2014a,Fredenhagen2013a}. Metrology techniques applied to quantum field theory in curved spacetime enables the estimation of spacetime parameters~\cite{Ahmadi2014b,Sabin2014a,Wang2014a,Shamsi2014a,Ahmadzadegan2014a,Yao2014quantum,Lindkvist2015a,Tian2015a,Friis2015a,kish2016estimating} using quantum systems providing a method that is compatible with both quantum and relativistic principles. Applications include accelerometers, gravitational wave detectors and relativistic quantum clocks. The parameters in this case are encoded through Bogoliubov transformations on initial quantum field states.

The paradigmatic predictions of the quantum field theory in curved spacetime are particle creation in an expanding Universe and Hawking radiation~\cite{Hawking1974a,Unruh1979a}. The theory has not been demonstrated in the laboratory yet and currently, experiments are aimed at testing analogue effects~\cite{Ade2014a,Karkare2014a,Belgiorno2010a}. However, recent results show that real space-time effects can be measured with current technologies. In particular, it has been shown that particle creation by gravitational waves produce observable effects in the phononic field of a Bose-Einstein condensate~\cite{Debs2011a,Bruschi2014a}. Space-time effects on BECs can be used to develop a new generation of measurement instruments by applying quantum metrology techniques. Until now, techniques in relativistic quantum metrology have focused on using pure states as quantum probes. However, in practice, systems interact with the environment and probe states are usually mixed. In this paper we compute explicit formulas to estimate parameters of quantum field theory in flat and curved space-time in the case that the probe states are in a thermal (mixed) state. Because of their mathematical simplicity, we restrict our analysis to Gaussian states.

The structure of the paper is the following. In section~\ref{sec:phase_space} we present the basic notions of quantum metrology for Gaussian states using the covariance matrix formalism. In section~\ref{sec:quantum_channels} we review the quantisation of relativistic fields. In section~\ref{sec:qfi} we compute the expression of the quantum Fisher information associated to one- and two-mode mixed Gaussian states as a function of the Bogoliubov coefficients. Then we compute explicitly the quantum Fisher information for the case where we wish to estimate a state parameter around the value $\epsilon_{0}\,=\,0$. Finally, we apply these results to calculate the quantum Fisher information for the estimation of the proper acceleration using one- and two-mode squeezed thermal states. We conclude with a review of our results and identify future directions.

Throughout the paper, matrices and vectors will be denoted in bold, $\boldsymbol{M},\boldsymbol{v}$. Identity matrices will be represented as $\boldsymbol{I}$ and we will be using Planck units $\hbar\,=\,c\,=\,k_{B}\,=\,1$.

\section{Phase space formalism of quantum metrology}\label{sec:phase_space}

In this section we outline the basic tools of quantum metrology for Bosonic systems via their quantum phase space description. This formalism considerably simplifies computations for continuous variable systems when the analysis is restricted to \emph{Gaussian states}. For a more detailed analysis of Gaussian states, we refer the reader to~\cite{Weedbrook2012a,Adesso2014a} and appendix~\ref{app1}.

We consider a system that consists of a collection of Bosonic modes described by creation and annihilation operators, $\hat{a}_{j},\hat{a}_{j}^{\dag}$. It is convenient to introduce a vector of operators $\boldsymbol{\hat{A}}\,=\,\boldsymbol{\hat{a}}\oplus\boldsymbol{\hat{a}}^{\dag}$, where $\boldsymbol{\hat{a}}\,=\,(\hat{a}_{1},\hat{a}_{2},\ldots)$ and the symbol $^\dag$ denotes Hermitian conjugation. The  canonical commutation relations are given by
\begin{equation}
[\hat{A}_{m},\hat{A}_{n}^{\dag}]\,=\,K_{mn}\,\mathrm{id}\quad\Rightarrow\quad\boldsymbol{K}\,=\,
\begin{bmatrix}
\boldsymbol{I} & \boldsymbol{0} \\
\boldsymbol{0} & -\boldsymbol{I}
\end{bmatrix},
\end{equation}
where $\mathrm{id}$ denotes the identity element of an algebra. Note that $\boldsymbol{K}^{-1}=\boldsymbol{K}^\dag=\boldsymbol{K}$ and $\boldsymbol{K}^2=\boldsymbol{I}$. The operators $\hat{A}_{j}$ are functions of the generalised position and momentum operators $\hat{x}_{j}$ and $\hat{p}_{j}$ (see appendix~\ref{app1}). These definitions define the ``complex form" of the continuous variable phase space~\cite{Arvind1995a}, which is particularly convenient when working with Bogoliubov transformations.

In quantum theory it is common to describe the state of the system using the density operator $\hat{\rho}$, which satisfies the normalisation condition $\mathrm{tr}\,\hat{\rho}\,=\,1$ where $\mathrm{tr}$ denotes the trace map. The density operator is positive semi-definite and self-adjoint in the Hilbert space inner product. An alternative, and completely equivalent, description of a bosonic quantum state is given by the \emph{symmetric characteristic function} defined as
\begin{equation}
\chi(\boldsymbol{\xi})\,=\,\mathrm{tr}\,[\hat{\rho}\,\hat{D}(\boldsymbol{\xi})],
\end{equation}
Note here that the operator $\hat{D}(\boldsymbol{\xi})\,=\,e^{\boldsymbol{\hat{A}}^{\dag}\boldsymbol{K}\boldsymbol{\xi}}$ is the \emph{Weyl displacement operator} with a complex variable of the form $\boldsymbol{\xi}\,=\,\boldsymbol{\gamma}\oplus\overline{\boldsymbol{\gamma}}$ which represents the displacement vector for the modes under consideration. Gaussian states are defined as states with characteristic functions of the form,
\begin{equation}
\chi(\boldsymbol{\xi})\,=\,e^{-\frac{1}{4}\boldsymbol{\xi}^{\dag}\boldsymbol{\sigma}\boldsymbol{\xi}-i\,\boldsymbol{d}^{\dag}\boldsymbol{K}\boldsymbol{\xi}}.
\end{equation}
 Gaussian states are completely represented by their first and second statistical moments, which can be collected in the vector $\boldsymbol{d}$ and covariance matrix $\boldsymbol{\sigma}$. The statistical moments are given in terms of the density operators
 $\hat{\rho}$ by,
\begin{subequations}
\begin{align}
\boldsymbol{d}&\,=\,\mathrm{tr}\,[\hat{\rho}\,\boldsymbol{\hat{A}}],\\
\boldsymbol{\sigma}&\,=\,\mathrm{tr}\,[\hat{\rho}\,\{\boldsymbol{\hat{A}},\boldsymbol{\hat{A}}^{\dag}\}].
\end{align}
\end{subequations}
The anti-commutator in the second moments is given ``element-wise" in the entires of $\boldsymbol{\hat{A}}$. In the convention used in this paper the vacuum state is given by the identity matrix $\boldsymbol{I}$, i.e. the variance of the quadrature operators $\hat{x}_{j}$ and $\hat{p}_{j}$ are $\mathrm{var}\,(\hat{x}_{j})\,=\,\mathrm{var}\,(\hat{p}_{j})\,=\,1$. Other work often defines the vacuum variances as $1/2$. In the complex form, the first and second moments of a state have block forms given by,
\begin{equation}
\boldsymbol{d}\,=\,
\begin{bmatrix}
\boldsymbol{d}_{\boldsymbol{a}} \\ \overline{\boldsymbol{d}}_{\boldsymbol{a}}
\end{bmatrix},\quad
\boldsymbol{\sigma}\,=\,\begin{bmatrix}
\boldsymbol{X} & \boldsymbol{Y} \\
\overline{\boldsymbol{Y}} & \overline{\boldsymbol{X}}
\end{bmatrix},
\end{equation}
where $\boldsymbol{X}^{\dag}\,=\,\boldsymbol{X}$ and $\boldsymbol{Y}^{\mathrm{tp}}\,=\,\boldsymbol{Y}$, and $\mathrm{tp}$ denotes transposition. For further details on the complex form of the first and second moments and their relation to the ``real" form that is often used in the literature see appendix~\ref{app1}. At the level of first and second moments, unitary transformations on the state are represented by \emph{symplectic} transformations and \emph{displacement vectors}. In particular, given a Gaussian preserving unitary $\hat{U}$, the first and second moments of a Gaussian state change as
\begin{subequations}
\label{eqn:moments_transformation}
\begin{align}
\boldsymbol{d}^{\prime}&\,=\,\boldsymbol{S}\,\boldsymbol{d}+\boldsymbol{b},\\
\boldsymbol{\sigma}^{\prime}&\,=\,\boldsymbol{S}\,\boldsymbol{\sigma}\,\boldsymbol{S}^{\dag},
\end{align}
\end{subequations}
and $\boldsymbol{b}$ denotes an arbitrary displacement in phase space.
The defining property of a symplectic matrix $\boldsymbol{S}$ is~\cite{Arvind1995a}
\begin{equation}
\boldsymbol{S}\,\boldsymbol{K}\,\boldsymbol{S}^{\dag}\,=\,\boldsymbol{K}.
\end{equation}
Some examples of Gaussian unitaries are the beam splitter and two-mode squeezing operators~\cite{Braunstein2005a}, which are well known operations in quantum optics.

A fundamental quantity in quantum information is the \emph{fidelity}, which quantifies the ``distance'' between two states. The fidelity typically employed in the literature is the Uhlmann fidelity~\cite{Jozsa1994a,Marian2012a}. Given two quantum states $\hat{\rho}_{1}$ and $\hat{\rho}_{2}$, the Uhlmann fidelity $\mathcal{F}$ takes the form
\begin{equation}
\mathcal{F}(\hat{\rho}_{1},\hat{\rho}_{2})\,=\,\bigg[\mathrm{tr}\sqrt{\sqrt{\hat{\rho}_{1}}\,\hat{\rho}_{2}\,\sqrt{\hat{\rho}_{1}}}\bigg]^{2}.
\end{equation}
The square roots above denote the \emph{unique} positive semi-definite operator which has the property $\sqrt{\hat{\rho}}\,\sqrt{\hat{\rho}}\,=\,\hat{\rho}$. Making the identification $\hat{\rho}_{1}\rightarrow(\boldsymbol{d}_{1},\boldsymbol{\sigma}_{1})$ and $\hat{\rho}_{2}\rightarrow(\boldsymbol{d}_{2},\boldsymbol{\sigma}_{2})$, we define the difference of the two displacement vectors as $\delta\boldsymbol{d}\,=\,\boldsymbol{d}_{1}-\boldsymbol{d}_{2}$ and the following useful quantities,
\begin{subequations}
\begin{align}
\Delta &\,\dot{=}\, \mathrm{det}\big[\boldsymbol{\sigma}_{1}+\boldsymbol{\sigma}_{2}\big], \\
\Gamma &\,\dot{=}\, \mathrm{det}\big[\boldsymbol{K}\,\boldsymbol{\sigma}_{1}\,\boldsymbol{K}\,\boldsymbol{\sigma}_{2}+\boldsymbol{I}\big],\\
\Lambda &\,\dot{=}\, \mathrm{det}\big[\boldsymbol{\sigma}_{1}+\boldsymbol{K}\big]\mathrm{det}\big[\boldsymbol{\sigma}_{2}+\boldsymbol{K}\big].
\end{align}
\end{subequations}
In terms of first and second moments, the fidelity between one- or two-mode Gaussian states then reads~\cite{Marian2012a},
\begin{subequations}
\label{eqn:QFI_defs_hilbert_space}
\begin{align}
\mathcal{F}_{1}(\hat{\rho}_{1},\hat{\rho}_{2})&\,=\,2\,\frac{e^{-\delta\boldsymbol{d}^{\dag}(\boldsymbol{\sigma}_{1}+\boldsymbol{\sigma}_{2})^{-1}\delta\boldsymbol{d}}}{\sqrt{\Delta+\Lambda}-\sqrt{\Lambda}}, \\
\mathcal{F}_{2}(\hat{\rho}_{1},\hat{\rho}_{2})&\,=\,4\,\frac{e^{-\delta\boldsymbol{d}^{\dag}(\boldsymbol{\sigma}_{1}+\boldsymbol{\sigma}_{2})^{-1}\delta\boldsymbol{d}}}{\sqrt{\Gamma}+\sqrt{\Lambda}-\sqrt{(\sqrt{\Gamma}+\sqrt{\Lambda})^{2}-\Delta}}.
\end{align}
\end{subequations}
Note that we have used different factors of 2 and the matrix $\boldsymbol{K}$ instead of the usual Gaussian symplectic form used by other authors. This is for a matter of convenience only.

We now define and compute the quantum Fisher information. There are usually two definitions for the quantum Fisher information. One is given by the \emph{Bures distance} induced by the Uhlmann fidelity on the space of quantum states, while the second is given by the \emph{symmetric logarithmic derivatives}~\cite{Holevo2011a,Petz2010a}. For our purposes, we will employ the Uhlmann fidelity for one and two mode systems.

The Bures distance $\mathrm{d}_{B}$ for two states $\hat{\rho}_{1},\hat{\rho}_{2}$ is defined using the Uhlmann fidelity~\cite{Paris2009a},
\begin{equation}
\mathrm{d}^{2}_{B}(\hat{\rho}_{1},\hat{\rho}_{2})\,=\,2\big(1-\sqrt{\mathcal{F}(\hat{\rho}_{1},\hat{\rho}_{2})}\big).
\end{equation}
One can then define the quantum Fisher information $H_{j}$ as~\cite{Hayashi2006a},
\begin{equation}
\label{eqn:QFI_def}
H_{j}(\epsilon)\,=\,\lim_{\mathrm{d}\epsilon\rightarrow 0}\,8\,\Bigg(\frac{1-\sqrt{\mathcal{F}_{j}(\hat{\rho}_{\epsilon},\hat{\rho}_{\epsilon+\mathrm{d}\epsilon})}}{\mathrm{d}\epsilon^{2}}\Bigg),
\end{equation}
where $\mathcal{F}_{j}$, $j=1,2$, denotes the $j$-mode Uhlmann fidelity between the state $\hat{\rho}_{\epsilon}$ and a state $\hat{\rho}_{\epsilon+\mathrm{d}\epsilon}$ infinitesimally separated away from it. The existence of the quantum Fisher information is dependent upon the differentiability of the Uhlmann fidelity around the point of interest $\epsilon$. Furthermore, the first derivative of the fidelity must vanish such that the limit in Eq.~\eqref{eqn:QFI_def} is well defined. Further details about this can be found in~\cite{Safranek2015a}.
Therefore, the one- and two-mode quantum Fisher information in terms of the first and second moments are given by,
\begin{equation}
\label{eqn:one_mode_quantum_fisher_information}
H_{1}(\epsilon)\,=\, \frac{1}{2}\frac{\mathrm{tr}\Big[\big(\boldsymbol{\Xi}^{-1}\dot{\boldsymbol{\Xi}}\big)^{2}\Big]}{1+\mathrm{det}[\boldsymbol{\Xi}]^{-1}}+\frac{1}{2}\,\frac{(\mathrm{det}[\boldsymbol{\Xi}])^{-1}\mathrm{tr}[\boldsymbol{\Xi}^{-1}\boldsymbol{\dot{\Xi}}]^{2}}{1-\mathrm{det}[\boldsymbol{\Xi}]^{-2}},
\end{equation}
\begin{widetext}
\begin{equation}
\begin{split}\label{eqn:two_mode_quantum_fisher_information}
H_{2}(\epsilon)\,&=\,\frac{1}{2(\mathrm{det}[\boldsymbol{\Xi}]-1)}\Bigg(\mathrm{det}[\boldsymbol{\Xi}]\mathrm{tr}\Big[\big(\boldsymbol{\Xi}^{-1}\dot{\boldsymbol{\Xi}}\big)^{2}\Big]+\sqrt{\mathrm{det}\big[\boldsymbol{I}+\boldsymbol{\Xi}^2\big]}\mathrm{tr}\Big[\big((\boldsymbol{I}+\boldsymbol{\Xi}^2)^{-1}\boldsymbol{\dot{\Xi}}\big)^2\Big]\\
&\quad+\frac{4(1+\mathrm{det}[\boldsymbol{\Xi}])(\mathrm{tr}[\boldsymbol{\Xi}\dot{\boldsymbol{\Xi}}]^{2}-\mathrm{det}[\boldsymbol{\Xi}]\mathrm{tr}[\boldsymbol{\Xi}^{-1}\dot{\boldsymbol{\Xi}}]^{2})+\mathrm{det}[\boldsymbol{\Xi}]\mathrm{tr}[\boldsymbol{\Xi}^{-1}\dot{\boldsymbol{\Xi}}]\mathrm{tr}[\boldsymbol{\Xi}^{2}](\mathrm{tr}[\boldsymbol{\Xi}^{-1}\dot{\boldsymbol{\Xi}}]\mathrm{tr}[\boldsymbol{\Xi}^{2}]-4\,\mathrm{tr}[\boldsymbol{\Xi}\dot{\boldsymbol{\Xi}}])}{4(1+\mathrm{det}[\boldsymbol{\Xi}])^{2}-\mathrm{tr}[\boldsymbol{\Xi}^{2}]^{2}}\Bigg).
\end{split}
\end{equation}
\end{widetext}
In the above $\boldsymbol{\Xi}\,\dot{=}\,\boldsymbol{K}\boldsymbol{\sigma}$ and $\dot{\boldsymbol{\Xi}}$ denotes the derivative with respect to the parameter $\epsilon$. Its is also possible to include the contribution to the quantum Fisher information from changes in displacement by adding the term $2\,\boldsymbol{\dot{d}}(\epsilon)^{\dag}\boldsymbol{\sigma}(\epsilon)^{-1}\boldsymbol{\dot{d}}(\epsilon)$ to either Eq.~\eqref{eqn:one_mode_quantum_fisher_information} or Eq.~\eqref{eqn:two_mode_quantum_fisher_information}.

The quantum Fisher information was derived in~\cite{Pinel2012a} in the single mode case and in~\cite{Safranek2015a} for two-mode states. In this paper we will consider that the first moments do not change with a small variation in epsilon. i.e. $\boldsymbol{\dot{d}}\,=\,\boldsymbol{0}$.

\section{Modeling quantum channels in quantum field theory}\label{sec:quantum_channels}

Quantum metrology provides strategies to estimate very precisely physical quantities related to quantum mechanical systems such as time, temperature and magnetic field strengths. However, estimating parameters that play a role in quantum field theory in flat and curved spacetime promise to enable the development of new measurement technologies such as gravimeters, accelerometers, relativistic quantum clocks and gravitational wave detectors~\cite{Ahmadi2014b,Lindkvist2015a,Sabin2014a}. In order to do so it is necessary to apply quantum metrology techniques to relativistic quantum fields and develop formulas for the quantum Fisher information in terms of Bogoliubov transformations.
In quantum field theory parameters are encoded into the state of the system through a channel that is implemented by a Bogoliubov transformation~\cite{Doukas2014a}. Such transformations arise when two different observers describe the same quantum field and when the spacetime undergoes some kind of dynamical transformation. More details can be found in~\cite{Benini2013a,Hollands2014a,Fredenhagen2013a,Birrell1984a,Ruijsenaars1978a}.

We consider a scalar field $\phi(t,\boldsymbol{x})$ that obeys the Klein-Gordon equation,
\begin{equation}\label{klein:gordon:equation}
\nabla^{\mu}\nabla_{\mu}\,\phi(t,\boldsymbol{x})\,=\,0.
\end{equation}
The operator $\nabla_{\mu}$ is the covariant derivative defined with respect to the metric tensor $\boldsymbol{g}$ in some suitable coordinates $(t,\boldsymbol{x})$. We restrict our analysis to spacetimes that admit  global or asymptotic time-like Killing vector fields since in this case it is posible to quantise the field. We denote the positive frequency mode solutions of the Klein-Gordon equation~\eqref{klein:gordon:equation} by $u_{n}(t,\boldsymbol{x})$, where $n\in\mathbb{N}$ is the mode number. We can collect these solutions, along with their complex conjugates (negative frequency solutions), in the vector $\boldsymbol{U}\,=\,\boldsymbol{u}\oplus\overline{\boldsymbol{u}}$ with $\boldsymbol{u}\,=\,(u_{1},u_{2},\ldots)$. The field is quantised by associating bosonic annihilation (creation) operators $a_{n}$ ($a^{\dagger}_{n}$) to the positive (negative) mode solutions of the Klein-Gordon equation,
\begin{equation}\label{field:expansion}
\phi(t,\boldsymbol{x})\,=\,\sum_{n}\big(a_{n}u_{n}(t,\boldsymbol{x})+\mathrm{c.c.}\big).
\end{equation}
The bosonic operators satisfy the canonical commutation relations,
\begin{equation}
\big[\hat{a}_{m},\hat{a}^{\dag}_{n}\big]\,=\,\delta_{mn}\,\mathrm{id}.
\end{equation}
In vector form the field is given by,
\begin{equation}
\phi(t,\boldsymbol{x})\,=\,\begin{bmatrix}
\boldsymbol{u} \\ \overline{\boldsymbol{u}}
\end{bmatrix}.\begin{bmatrix}
\boldsymbol{a} \\ \overline{\boldsymbol{a}}
\end{bmatrix}
\end{equation}
where the dot product is defined as $\boldsymbol{x}\cdot\boldsymbol{y}\,\dot{=}\,\boldsymbol{x}^{\mathrm{tp}}\boldsymbol{y}$, for vectors $\boldsymbol{x}$, $\boldsymbol{y}$. The vacuum state of the field $|0\rangle$ is defined by the equation $\hat{a}_{n}|0\rangle\,=\,0$ for all $n$.

The field expansion~\eqref{field:expansion} is not unique. It can be written in a different basis of solutions to the Klein-Gordon equation denoted $v_{n}(t,\boldsymbol{x})$ and in vector form given by $\boldsymbol{V}\,=\,\boldsymbol{v}\oplus\overline{\boldsymbol{v}}$. The space of solutions is a linear vector space. Therefore, the solutions are related via, \begin{equation}
\label{eqn:phase_space_linear_transformation}
\boldsymbol{U}\,=\,\boldsymbol{S}\,\boldsymbol{V}
\end{equation}
where the matrix $\boldsymbol{S}$ is called a Bogoliubov transformation. It is easy to show that the transformation has the following block structure,
\begin{equation}
\label{eqn:bogoliubov_transformation_definition}
\begin{bmatrix}
\boldsymbol{u} \\ \overline{\boldsymbol{u}}
\end{bmatrix}\,=\,
\begin{bmatrix}
\boldsymbol{\alpha} & \boldsymbol{\beta} \\
\overline{\boldsymbol{\beta}} & \overline{\boldsymbol{\alpha}}
\end{bmatrix}\,
\begin{bmatrix}
\boldsymbol{v} \\ \overline{\boldsymbol{v}}
\end{bmatrix}.
\end{equation}
where $\boldsymbol{\alpha}$ and $\boldsymbol{\beta}$ are the Bogoliubov coefficients \cite{Birrell1984a}. The choice of basis $\boldsymbol{U}$ or $\boldsymbol{V}$ is equivalent therefore,
\begin{equation}
\hat{\phi}\,(t,\boldsymbol{x})\,=\,\begin{bmatrix}
\boldsymbol{u} \\ \overline{\boldsymbol{u}}
\end{bmatrix}.\begin{bmatrix}
\boldsymbol{\hat{a}} \\ \boldsymbol{\hat{a}}^{\dag}
\end{bmatrix}\,\equiv\,\begin{bmatrix}
\boldsymbol{v} \\ \overline{\boldsymbol{v}}
\end{bmatrix}.\begin{bmatrix}
\boldsymbol{\hat{b}} \\ \boldsymbol{\hat{b}}^{\dag}
\end{bmatrix}.
\end{equation}
This equivalence defines the dual of Eq.~\eqref{eqn:bogoliubov_transformation_definition} for the transformation between different annihilation and creation operators,
\begin{equation}
\label{eqn:operator_bogoliubov_transformation_definition}
\begin{bmatrix}
\boldsymbol{\hat{a}} \\ \boldsymbol{\hat{a}}^{\dag}
\end{bmatrix}\,=\,
\begin{bmatrix}
\overline{\boldsymbol{\alpha}} & -\overline{\boldsymbol{\beta}} \\
-\boldsymbol{\beta} & \boldsymbol{\alpha}
\end{bmatrix}\,
\begin{bmatrix}
\boldsymbol{\hat{b}} \\ \boldsymbol{\hat{b}}^{\dag}
\end{bmatrix}.
\end{equation}
The transformation between the real and complex form of the Bogoliubov transformations can be found in appendix~\ref{app2}. As the Bogoliubov transformation is a linear transformation on the classical phase space, it should not influence the canonical quantisation of the field and hence should preserve the canonical commutation relations of the mode operators. This is also a statement of the preservation of commutation relations under unitary transformations. This requirement leads to the well known Bogoliubov identities
\begin{subequations}
\label{eqn:bogo_identities}
\begin{align}
\boldsymbol{\alpha}\,\boldsymbol{\alpha}^{\dag}-\boldsymbol{\beta}\,\boldsymbol{\beta}^{\dag}&\,=\, \boldsymbol{I}, \\
\boldsymbol{\alpha}\,\boldsymbol{\beta}^{\mathrm{tp}}-\boldsymbol{\beta}\,\boldsymbol{\alpha}^{\mathrm{tp}}&\,=\,0.
\end{align}
\end{subequations}
The Bogoliubov identities \eqref{eqn:bogo_identities} imply that the Bogoliubov transformation $\boldsymbol{S}$ satisfies
\begin{equation}
\label{eqn:symplectic_definition}
\boldsymbol{S}\,\boldsymbol{K}\,\boldsymbol{S}^{\dag}\,=\,\boldsymbol{K}.
\end{equation}
This condition coincides with the definition of the symplectic group, as identified in the previous section. We notice that due to the infinite dimensional nature of the field basis, the matrices $\boldsymbol{S}$ are infinite dimensional. This implies that all vector or matrix expressions are to be understood element-wise. With this in mind, the mode operator transformation in Eq.~(\ref{eqn:operator_bogoliubov_transformation_definition}) can be written as
\begin{equation}
\label{eqn:operator_bogoliubov_transformation}
\boldsymbol{\tilde{S}}\,\dot{=}\,\begin{bmatrix}
\overline{\boldsymbol{\alpha}} & -\overline{\boldsymbol{\beta}} \\
-\boldsymbol{\beta} & \boldsymbol{\alpha}
\end{bmatrix}\,=\big(\boldsymbol{S}^{-1}\big)^{\mathrm{tp}}\,=\,\boldsymbol{K}\,\overline{\boldsymbol{S}}\,\boldsymbol{K},
\end{equation}
which is also symplectic.

We can use the inner product of the Klein-Gordon solutions to calculate the transformation coefficients as~\cite{Crispino2008}
\begin{subequations}
\begin{align}
\alpha_{mn}&\,=\, \big(u_{m},v_{n}\big)\big|_{\Sigma},\\
\beta_{mn}&\,=\, \big(u_{m},\overline{v}_{n}\big)\big|_{\Sigma}.
\end{align}
\end{subequations}
These coefficients encode the information of a transformation between two sets of solutions of the Klein-Gordon equation on a given time-like hypersurface $\Sigma$. However, as they are defined only for a fixed time, they will not in general be suitable to describe the continuous evolution of a quantum state. In particular, our Bogoliubov transformations transform an initial state at time $\tau_{0}$ to a final state at time $\tau$. These general Bogoliubov transformations will therefore depend on $\epsilon$ (our parameter of interest) and the time elapsed between initial and final state defined as $\tau-\tau_{0}$. As our results do not depend on the particular method of construction, we only require they satisfy the Bogoliubov identities in Eq.~\eqref{eqn:bogo_identities}.  Hence we shall neglect the technical details of continuous Bogoliubov transformation construction. For a detailed account of constructing various types of continuous Bogoliubov transformations see~\cite{Bruschi2013b}.

The transformation between the initial and final state of the field, when restricted to the Gaussian case, is given by the Bogoliubov matrix $\tilde{\boldsymbol{S}}$ and by the transformation properties of the first and second field moments Eq~(\ref{eqn:moments_transformation}). We consider $\tilde{\boldsymbol{\sigma}}_{0}$ to be the initial covariance matrix of the field and  $\mathrm{tr}_{E}[\tilde{\boldsymbol{\sigma}}_{0}]\,=\,\boldsymbol{\sigma}_{0}$ is the (one or two-mode) reduced state of interest. We assume that $\mathrm{tr}_{\neg E}[\tilde{\boldsymbol{\sigma}}_{0}]\,=\,\boldsymbol{\sigma}_{E}$ are the remaining modes (or ``environment'' $E$) which contains no initial correlation with the system modes $\boldsymbol{\sigma}_{0}$. Therefore, the initial state $\tilde{\boldsymbol{\sigma}}_{0}$ is separable in the subsystem-environment bipartition. Concretely, the block structure of the initial state is,
\begin{equation}
\tilde{\boldsymbol{\sigma}}_{0}\,\dot{=}
\begin{bmatrix}
\boldsymbol{X}_{0} & \boldsymbol{Y}_{0} \\
\overline{\boldsymbol{Y}}_{0} & \overline{\boldsymbol{X}}_{0}
\end{bmatrix}
\end{equation}
Next, we can describe the transformation from the initial subsystem state $\boldsymbol{\sigma}_{0}$ to a final subsystem state $\boldsymbol{\sigma}(\epsilon)$ via a map $\mathcal{E}$ defined as
\begin{equation}
\label{eqn:QFI_quantum_channel}
\mathcal{E}[\boldsymbol{\sigma}_{0}]\,=\,\mathrm{tr}_{E}\big[\boldsymbol{\tilde{S}}(\epsilon)\,\tilde{\boldsymbol{\sigma}}_{0}\,\boldsymbol{\tilde{S}}(\epsilon)^{\dag}\big]\,\dot{=}\,\boldsymbol{\sigma}(\epsilon).
\end{equation}
The Bogoliubov transformation between the initial and final states can be viewed as a quantum channel on the space of quantum states. The expression for an arbitrary Gaussian state, and in particular, for initial one- and two-mode squeezed thermal states, can be found in appendix~\ref{app3}. As an example, we choose a state initially in the vacuum $\tilde{\boldsymbol{\sigma}}_{0}\,=\,\boldsymbol{I}$ given by $X_{0,ab}\,=\,\delta_{ab}$ and $Y_{0,ab}\,=\,0$.  Therefore, the reduced state after a Bogoliubov  transformation takes the form,
\begin{subequations}
\begin{align}
X_{ij}&\,=\,\sum_{a}\Big(\alpha_{ia}\,\overline{\alpha}_{ja}+\beta_{ia}\overline{\beta}_{ja}\Big),\\
Y_{ij}&\,=\,\sum_{a}\Big(\beta_{ia}\,\alpha_{ja}+\alpha_{ia}\,\beta_{ja}\Big).
\end{align}
\end{subequations}

\section{Ultimate precision: Quantum Fisher Information for bosonic fields}\label{sec:qfi}

We now compute the quantum Fisher information for Bogoliubov channels that encode physical quantities of interest in quantum field theory in curved space-time. We will pay particular attention to the case where we want to estimate the difference between an initial state undergoing some transformation and a state that does not undergo the transformation. This is equivalent to evaluating the quantum Fisher information at $\epsilon_{0}\,=\,0$. We will also implicitly assume that the Bogoliubov coefficients have the following property
\begin{equation}
\label{eqn:simplifying_condition}
\frac{\mathrm{d}}{\mathrm{d}\epsilon}\alpha_{jj}(\epsilon)\big|_{\epsilon\,=\,0}\,=\,\frac{\mathrm{d}}{\mathrm{d}\epsilon}\beta_{jj}(\epsilon)\big|_{\epsilon\,=\,0}\,=\,0.
\end{equation}
This is equivalent to the statement that the first order coefficients of the diagonal $\boldsymbol{\alpha}$ and $\boldsymbol{\beta}$ are zero i.e. $\alpha^{(1)}_{jj}\,=\,\beta^{(1)}_{jj}\,=\,0$. As an example, these assumptions hold when the symplectic operation is symmetric around zero, i.e., $\alpha(\epsilon)=\alpha(-\epsilon)$, $\beta(\epsilon)=\beta(-\epsilon)$, and also for the special case of $\alpha_{mn},\beta_{mn}\in\mathbb{R}$. Physically, this condition means that the channel does not affect the same mode up to the first order in $\epsilon$. I.e., the channel is mostly mode-entangling channel. It is possible to generalise this work to cases where diagonal first order Bogoliubov coefficients are non-zero, however, since Bogoliubov coefficients considered in previous literature~\cite{Bruschi2012a,Bruschi2013b,Friis2013a,Sabin2014a} all satisfy Eq.~\eqref{eqn:simplifying_condition}, we will restrict to such case. In the following sections, we consider that all quantities (matrix and scalar) can be expanded in the form,
\begin{equation}
f(\epsilon)\,=\,f^{(0)}+f^{(1)}\epsilon+f^{(2)}\epsilon^{2}+\mathcal{O}(\epsilon^{3}).
\end{equation}

\subsection{One-mode systems}

We first compute the quantum Fisher information of a single mode undergoing a Bogoliubov transformation that depends on the physical parameter to be estimated. We consider the following initial state, \begin{equation}
\label{eqn:Single_mode_state}
\boldsymbol{\sigma}_{0}\,=\,\nu_{m}\,
\begin{bmatrix}
\cosh(2r) & \sinh(2r) \\
\sinh(2r) & \cosh(2r)
\end{bmatrix},\quad\boldsymbol{\sigma}_{E}\,=\,\bigoplus_{j\ne m}\nu_{j}\,\boldsymbol{I},
\end{equation}
which corresponds to a single mode squeezed thermal state with squeezing parameter $r$, thermal parameter $\nu_{m}\ge 1$ and all other modes in a separable thermal state. The temperature of the state, denoted by $T$, is related to the thermal parameter through $\nu_{m}\,=\,\coth(E_{m}/2T)$ where $E_{m}=\omega_{m}$ is the energy of each mode. Note that for zero temperature, the thermal parameter reduces to $\nu_{m}\,=\,1$.

We can write the final state as a series expansion in $\epsilon$ around the point $\epsilon_{0}\,=\,0$,
\begin{equation}
\boldsymbol{\sigma}(\epsilon)\,=\,\boldsymbol{\sigma}^{(0)}+\boldsymbol{\sigma}^{(1)}\,\epsilon+\mathcal{O}(\epsilon^{2}).
\end{equation}
The exact final state elements can be computed using the expressions in appendix~\ref{eqn:one_mode_XYsummed}. It should also be noted that, in general, the covariance matrix elements $X_{mn}^{(j)}$ and $X_{mn}^{(j)}$ will depend on both squeezing, $r$, and the thermal parameters $\nu_{m}$. We will also denote phases acquired due to free time evolution as $G_{m}\,=\,e^{+i\,\omega_{m}\tau}$ with $\omega_{m}$ the zeroth order contribution to a modes frequency.

We now proceed to choose specific values for the temperature and squeezing to find analytically the quantum Fisher information in regimes of interest.

\subsubsection{Initial zero temperature}

We start by considering an initial state with zero temperature. The perturbative expansion of $\Lambda$  in Eqs.~\eqref{eqn:QFI_defs_hilbert_space} needs particular attention. If we consider a state which is initially pure one finds that the denominators in Eqs.~\eqref{eqn:one_mode_quantum_fisher_information} and~\eqref{eqn:two_mode_quantum_fisher_information} vanish. This potentially problematic point can be handled in multiple ways, for more details see~\cite{Safranek2015a}. However, one can make a series expansion of each term and by applying L'H\^{o}pital's rule one obtains a finite result,
\begin{widetext}
\begin{equation}
\label{eqn:one_mode_qfi_zero_temp}
H_{1}(\epsilon)\,=\,X_{mm}^{(2)}\,\cosh(2\,r)-\mathrm{Re}[G_{m}^{2}\,Y_{mm}^{(2)}]\,\sinh(2r)+\frac{2}{3}\Big(X_{mm}^{(3)}\,\cosh(2\,r)-\mathrm{Re}[G_{m}^{2}\,Y_{mm}^{(3)}]\,\sinh(2\,r)\Big)\,\epsilon+\mathcal{O}(\epsilon^{3})
\end{equation}
\end{widetext}
For convenience and clarity, we have left the second order covariance matrix elements written in the general form $X^{(2)}_{mm}$ and $Y^{(2)}_{mm}$. This elegant expression builds upon the zeroth order result of~\cite{Ahmadi2014a} and extends it to the linear regime in $\epsilon$. It should be noted that, for the expansion \eqref{eqn:one_mode_qfi_zero_temp} to be valid, the initial squeezing $r$ and parameter $\epsilon$ must satisfy $e^{2r}\,\epsilon\ll1$.

\subsubsection{Initial ``small" temperature}

A case of physical relevance is that of small temperature. In realistic situations the field is never in the vacuum state.  It is possible to compute analytical formulas in different regimes of interest that depend on the relative magnitude of the temperature parameter and the parameter $\epsilon$. We start by analysing the case where the temperature is ``small" as compared to $\epsilon$. The thermal parameter $\nu_{m}$ in this case has the form $\nu_{m}\,=\,1+2\,Z^{2}+\mathcal{O}(Z^{3})$, with $Z\,=\,e^{-E_{m}/2T}$. We can identify the ratio $Z^{2}/\epsilon^{2}$ as our new expansion parameter. This defines our ``small'' temperature regime as the one where $Z^{2}\ll\epsilon^{2}$. We find the quantum Fisher information takes the expression
\begin{equation}
\label{eqn:small_temperature_one_mode_qfi}
H_1(\epsilon)\,\approx\,H_{1}^{(0)}-4\frac{Z^{2}}{\epsilon^{2}}.
\end{equation}
One notes that the main contribution is the zeroth order and zero temperature QFI $H_{1}^{(0)}$ which coincides with the zeroth order contribution from Eq.~\eqref{eqn:one_mode_qfi_zero_temp}. The second term is a small correction.  We have neglected the linear contribution to the zero temperature result for simplicity. 

\subsubsection{Initial ``large" temperature}

In this case we find that the zeroth order quantum Fisher information for a single mode is identically zero, i.e. $H^{(0)}_{1}\,=\,0$ for any non-zero temperature. This implies that the estimation of the parameter $\epsilon$ around zero is impossible for a one-mode squeezed state with a non-zero temperature. The first non-trivial contribution comes at $\mathcal{O}(\epsilon^{2})$. The result is,
\begin{widetext}
\begin{equation}\label{one:mode:large:temperature:result}
H^{(2)}_{1}\,=\,\Bigg(\frac{|Y_{mm}^{(2)}|^{2}}{\nu_{m}^{2}+1}+\frac{(X_{mm}^{(2)})^{2}}{\nu_{m}^{2}-1}\,-\, \frac{2\,\nu_{m}^{2}\,X_{mm}^{(2)}\mathrm{Re}[G_{m}^{2}\overline{Y}_{mm}^{(2)}]}{\nu_{m}^{4}-1}\sinh(4r)\,+\, \frac{2\,\nu_{m}^{2}\Big((X_{mm}^{(2)})^{2}+\mathrm{Re}[G_{m}^{4}(\overline{Y}_{mm}^{(2)})^{2}]\Big)}{\nu_{m}^{4}-1}\sinh^{2}(2r)\Bigg)
\end{equation}
\end{widetext}
Clearly the condition $\nu_{m}>1$ is key and the equation holds in the regime $\epsilon^{2}\ll\nu_{m}-1$ or, in terms of the previous subsections notation, $\epsilon^{2}\ll Z^{2}$ by which the ``large" temperature regime is defined.

\subsection{Two-mode states}

Here we compute the quantum Fisher information for thermal two-mode states with non-degenerate thermal parameters (i.e., the frequencies of the two modes are different). The initial state has the form,
\begin{equation}
\label{eqn:Two_mode_state}
\boldsymbol{\sigma}_{0}\,=\,
\begin{bmatrix}
D_{mn} & 0 & 0 & C_{mn} \\
0 & D_{nm} & C_{mn} & 0 \\
0 & C_{mn} & D_{mn} & 0 \\
C_{mn} & 0 & 0 & D_{nm}`
\end{bmatrix},\,\,\boldsymbol{\sigma}_{E}\,=\,\bigoplus_{j\ne m,n}\nu_{j}\,\boldsymbol{I},
\end{equation}
where we have introduced
\begin{subequations}
\begin{align}
D_{mn}&\,\dot{=}\,\nu_{m}\cosh^{2}(r)+\nu_{n}\sinh^{2}(r), \\
C_{mn}&\,\dot{=}\,(\nu_{m}+\nu_{n})\cosh(r)\sinh(r).
\end{align}
\end{subequations}

\subsubsection{Initial zero temperature}

In the two-mode case, the QFI for zero temperature is given as a series expansion in $\epsilon$. The resulting expressions are computable but considerably more involved. Here we present the results for the zero and first order contributions in $\epsilon$. In the linear contribution, we present only the case of zero squeezing. The formula for non-zero squeezing is too long and therefore, we have chosen to focus on the quantitative behaviour.
\begin{subequations}
\begin{align}
\label{eqn:two_mode_qfi_zero_temperature_zeroth_order}
H_{2}^{(0)}\,=&\,(X_{mm}^{(2)}+X_{nn}^{(2)})\cosh(2r)\nonumber\\
&\,-4|\beta_{mn}^{(1)}|^{2}-2\,\mathrm{Re}[G_{m}G_{n}Y_{mn}^{(2)}]\sinh(2r)\nonumber\\
&\,-4\,(|\alpha_{mn}^{(1)}|^{2}+\mathrm{Im}[G_{m}\,\overline{\beta}_{mn}^{(1)}]^{2})\sinh^{2}(2r), \\
\label{eqn:two_mode_qfi_zero_temperature_first_order}
H_{2}^{(1)}\big|_{r\,=\,0}\,=\,&\frac{2}{3}\,\Big(6\,\mathrm{Re}[G_{n}\beta_{mn}^{(1)}Y_{mn}^{(2)}]+X_{mm}^{(3)}+X_{nn}^{(3)}]\Big).
\end{align}
\end{subequations}
At zeroth order, the quantum Fisher information depends on the squeezing parameter in the same way as in the single mode channels studied in the previous section. However, at first order, particle creation terms  $\beta_{mn}^{(1)}$ appear generating entanglement in the system. Therefore, we conclude that in this case entanglement does not provide an important improvement in precision. A highly squeezed single mode probe state could be enough to enable a good measurement strategy for the estimation of parameters uncoded in Bogoliubov transformations. Single mode states are usually more accessible in realistic experiments and this could provide an important advantage in quantum metrology for quantum fields.

\subsubsection{Initial ``small" temperature}

We now compute the ``small" temperature contribution to the QFI for two mode probe states by expanding the thermal parameters as $\nu_{m}\,=\,1+2\,q_{m}Z^{2}+\mathcal{O}(Z^{3})$ where $q_{m}$ is a fixed constant. We find that for zero squeezing, the first contribution to the QFI in the small temperature regimes has the following form,
\begin{equation}
\label{eqn:two_mode_small_temperature}
H_2(\epsilon)\,\approx\, H_{2}^{(0)}-4\,(q_{m}+q_{n})\frac{Z^{2}}{\epsilon^{2}},
\end{equation}
where the zeroth order $H_{2}^{(0)}$ coincides with Eq.~\eqref{eqn:two_mode_qfi_zero_temperature_zeroth_order}. It should be noted that due to the complexity of the two-mode non-zero squeezing expression, the first correction term in Eq.~\eqref{eqn:two_mode_small_temperature} has only been verified for zero squeezing.

\subsubsection{``Large" temperature}

As in the single mode case, the thermal parameters take values strictly greater than unity, i.e. $\nu_{j}>1$. The components of the state can be exactly computed and are given in appendix~\ref{eqn:two_mode_XYsummed}. We find that the quantum Fisher information, including linear contributions, is given by $H_2(\epsilon)\,=\,H^{(0)}_{2}+H^{(1)}_{2}\epsilon+\mathcal{O}(\epsilon^{2})$, with coefficients,
\begin{widetext}
\begin{equation}
\label{eqn:two_mode_QFI_zeroth_order}
\begin{split}
H_{2}^{(0)}&\,=\,h_{00}+h_{02}\,\sinh^{2}(2r), \\
h_{00}&\,=\,\frac{2\,(\nu_{m}-\nu_{n})^{2}|\alpha_{mn}^{(1)}|^{2}}{\nu_{m}\nu_{n}-1}+\frac{2\,(\nu_{m}+\nu_{n})^{2}|\beta_{mn}^{(1)}|^{2}}{\nu_{m}\nu_{n}+1}, \\
h_{02}&\,=\,\frac{2\,(\nu_{m}+\nu_{n})^{2}((\nu_{m}\nu_{n}-1)^{2}+\nu_{m}^{2}+\nu_{n}^{2}-2)|\alpha_{mn}^{(1)}|^{2}}{(\nu_{m}^{2}+1)(\nu_{n}^{2}+1)(\nu_{m}\nu_{n}-1)}+\frac{2\,(\nu_{m}+\nu_{n})^{2}\mathrm{Im}[G_{m}\overline{\beta}_{mn}^{(1)}]^{2}}{\nu_{m}\nu_{n}+1}.
\end{split}
\end{equation}
\begin{equation}
\begin{split}
H_{2}^{(1)}&\,=\,h_{10}+h_{11}\,\sinh(2r)+h_{12}\,\sinh^{2}(2r), \\
h_{10}&\,=\,4\,\Bigg(\frac{\nu_{n}-\nu_{m}}{\nu_{m}\nu_{n}-1}\,\mathrm{Re}[G_{n}\overline{\alpha}_{mn}^{(1)} \overline{X}_{mn}^{(2)}]-\frac{\nu_{m}+\nu_{n}}{\nu_{m}\nu_{n}+1}\,\mathrm{Re}[G_{n}\beta_{mn}^{(1)}Y_{mn}^{(2)}]\,\cosh(2r)\Bigg), \\
h_{11}&\,=\, \frac{2\,(\nu_{m}+\nu_{n})\mathrm{Re}[G_{m}\overline{\beta}_{mn}^{(1)}]}{\nu_{m}\nu_{n}+1}(X_{mm}^{(2)}+X_{nn}^{(2)})-\frac{16\,\nu_{m}\nu_{n}(\nu_{m}^{2}-\nu_{n}^{2})^{2}|\alpha_{mn}^{(1)}|^{2}\mathrm{Re}[G_{m}\overline{\beta}_{mn}^{(1)}]}{(\nu_{m}^{2}+1)(\nu_{n}^{2}+1)(\nu_{m}^{2}\nu_{n}^{2}-1)}\cosh(2r) \\
&\quad\quad-\frac{2(\nu_{m}+\nu_{n})(\nu_{m}\nu_{n}+1)}{(\nu_{m}^{2}+1)(\nu_{m}^{2}+1)}\mathrm{Re}[G_{n}\overline{\alpha}_{mn}^{(1)}(\overline{G}_{m}\overline{G}_{n} \overline{Y}_{mm}^{(2)}-G_{m}G_{n}Y_{nn}^{(2)})]\\
&\quad\quad+\frac{2(\nu_{m}-\nu_{n})(\nu_{m}+\nu_{n})^{2}}{(\nu_{m}^{2}+1)(\nu_{n}^{2}+1)(\nu_{m}\nu_{n}-1)}\mathrm{Re}[G_{n}\overline{\alpha}_{mn}^{(1)}(\overline{G}_{m}\overline{G}_{n} \overline{Y}_{mm}^{(2)}+G_{m}G_{n}Y_{nn}^{(2)})]\cosh(2r),\\
h_{12}&\,=\,\frac{4\,(\nu_{n}-\nu_{m})(\nu_{n}+\nu_{m})^{2}\mathrm{Re}[G_{n}\overline{\alpha}_{mn}^{(1)}\,\overline{X}_{mn}^{(2)}]}{(1+\nu_{m}^{2})(1+\nu_{n}^{2})(\nu_{m}\nu_{n}-1)}.
\end{split}
\end{equation}
\end{widetext}
In general, the coefficients $\alpha^{(1)}_{mn}$,  $\beta^{(1)}_{mn}$ and the covariance matrix are time dependent. In the large temperature regime, to zeroth order, the quantum Fisher information for two-mode probe states is non-zero. This result is in contrast with the single-mode case. Even when the probe state  has zero squeezing, the quantum Fisher information is non-zero and it is proportional to the number of particles created after the Bogoliubov transformation~\cite{Bruschi2013c}. We can also analyse the effect of temperature on the quantum Fisher information by varying the parameters $\nu_{m}$ and $\nu_{n}$.
We note that at when estimating around the point $\epsilon\,=\,0$, the zero order expressions for the quantum Fisher information are exact, $H(0)=H^{(0)}$.

\section{Example: Estimation of the proper acceleration}
\begin{figure}[t!]
\centering
\includegraphics[width=1\linewidth]{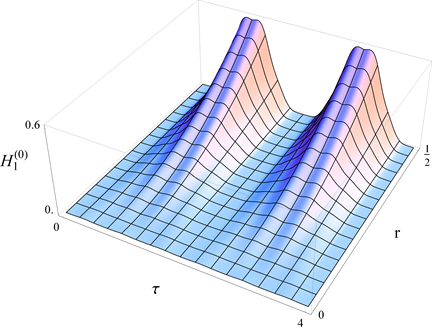}
\caption{The zeroth order of the quantum Fisher information for the estimation of the acceleration parameter $a$ using a one-mode squeezed state with initial zero temperature. Calculated using Eq.~\eqref{eqn:one_mode_qfi_zero_temp} while choosing $m=1$ (using a Fock space corresponding to the first excited state within a cavity). The graph shows that to achieve the highest possible precision in estimation it is appropriate to measure at certain times ($\tau=1,3,5,\dots$). This periodic behavior is due to the fact that the information about the parameter moves into the modes we cannot access -- the environment -- and back. At times when the quantum Fisher information is the highest the estimation precision grows exponentially with the squeezing parameter $r$.}\label{fig:one-modeQFI}
\end{figure}

\begin{figure}[t!]
\centering
\includegraphics[width=1\linewidth]{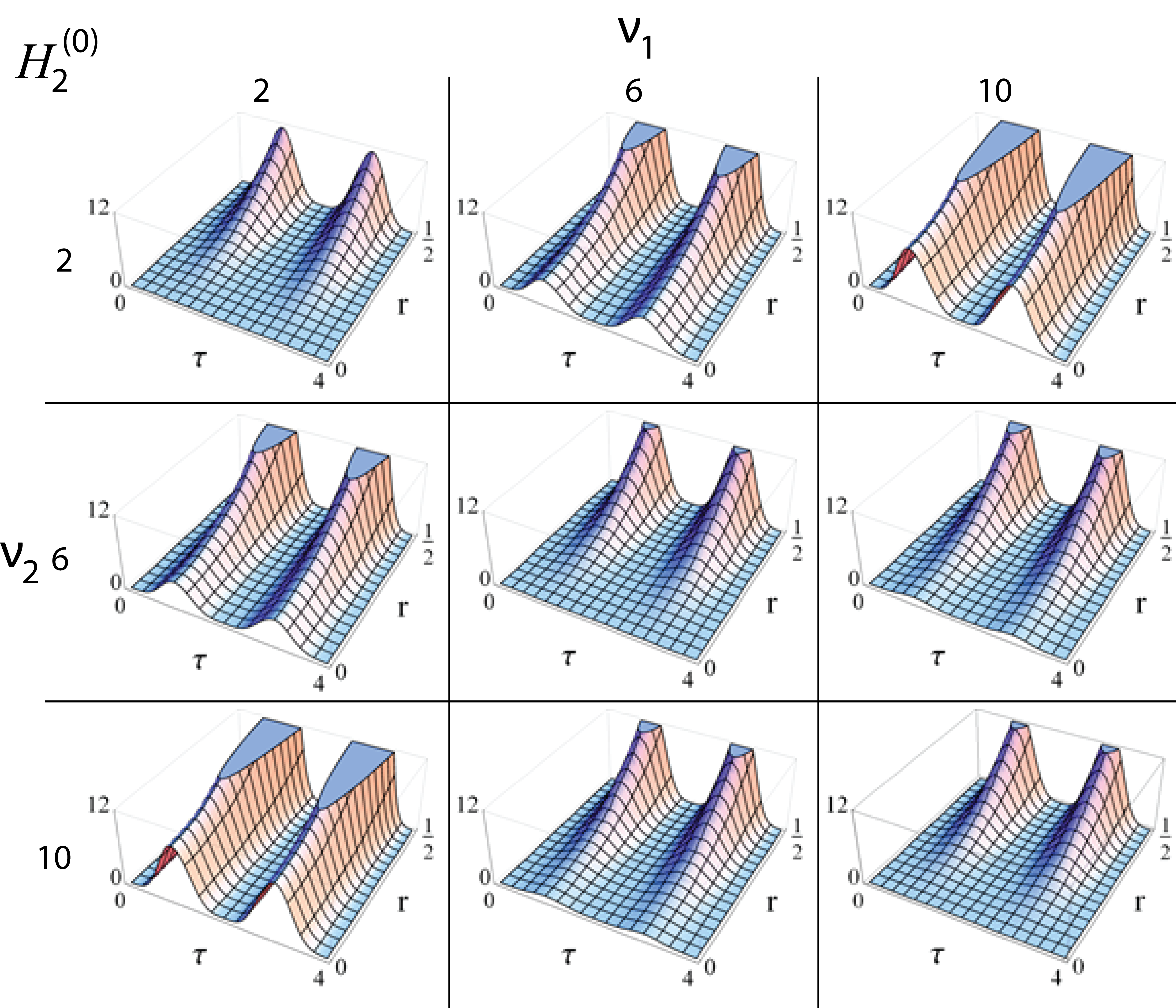}
\caption{The zeroth order of the quantum Fisher information for the estimation of the acceleration parameter $a$ using a two-mode squeezed thermal state with initial ``large'' temperature. Calculated using Eq.~\eqref{eqn:two_mode_QFI_zeroth_order} while choosing $m=1$, $n=2$ (Fock spaces corresponding to the first and the second excited state within the cavity). Different combinations of initial temperatures are used, $\nu_{m,n}=2,6,10$. Similarly to the one-mode scenario, it is appropriate to measure at certain times ($\tau=1,3,5,\dots$) when the estimation precision grows exponentially with the squeezing parameter $r$. Moreover, the graph shows that the highest precision in estimation is achieved with large temperature difference between the modes, i.e., when $\nu_1=2$ and $\nu_2=10$, or $\nu_1=10$ and $\nu_2=2$. The diagonal $\nu_1=\nu_2\rightarrow\infty$ quickly converges to the double of the two-mode squeezed vacuum value given by $\nu_1=\nu_2=1$. An opportunity of using temperature difference between the modes is not the only advantage of using the two-mode states. In contrast to the one-mode states, two-mode states also achieve one order higher precision with the same amount of squeezing.}\label{fig:two-modeQFI}
\end{figure}

To illustrate the power of the derived formulae, we calculate the bound on the estimation of the proper acceleration using cavities. Assume a quantum state inside of a non-moving cavity. Starting at proper time $\tau_0=0$, the cavity goes through a period $\tau$ of the proper acceleration $a$ (as measured in the centre of the cavity) and period $\tau$ of retardation $-a$, stopping again at time $2\tau$. Since we wish to estimate the proper acceleration $a$ we identify $a\equiv \epsilon$. The proper length of the cavity $L=1$ is considered constant during the whole procedure. Bogoliubov transformation of the state in this scenario has been calculated in~\cite{Bruschi2013b} using a continuous perturbative expansion in the small parameter $a$,
\begin{widetext}
\begin{subequations}
\begin{align}
\alpha_{mn}(a)&\,=\, e^{i\omega_n2\tau}+\mathcal{O}(a^2),\ &m=n\nonumber\\
&\,=\, -\frac{8i\sqrt{mn}}{(m-n)^3\pi^2}e^{\frac{1}{2}i\pi(m+n-2m\tau+6n\tau)}\sin{\tfrac{(m+n)\pi}{2}}\sin^2{\tfrac{(m-n)\pi\tau}{2}}\,a+\mathcal{O}(a^2),\ &m\neq n\\
\beta_{mn}(a)&\,=\, -\frac{8i\sqrt{mn}}{(m+n)^3\pi^2}e^{\frac{1}{2}i\pi(m+n-2m\tau+2n\tau)}\sin{\tfrac{(m+n)\pi}{2}}\sin^2{\tfrac{(m+n)\pi\tau}{2}}\,a+\mathcal{O}(a^2),&
\end{align}
\end{subequations}
\end{widetext}
where $\omega_n=\frac{n\pi}{L}$ is a natural frequency of the $n$-th mode. Using these transformations, we calculate the zeroth order quantum Fisher information for a one-mode squeezed vacuum and a two-mode squeezed thermal state as shown on figures~\ref{fig:one-modeQFI} and \ref{fig:two-modeQFI}.

\section{Conclusions}\label{sec:conclusions}

In order to provide a general framework for estimating parameters of interest in gravity and relativity using quantum metrology, we have extended previous pure state analysis to the mixed case. The main motivation is that, for any practical and experimental purposes, quantum systems are always mixed. We have restricted our analysis to Gaussian probe states since, in this case, the covariance matrix formalism provides advantageous simplifications of the mathematical description of the states. In particular, Gaussian states are also relatively easy to prepare in quantum optical laboratories. We have computed general and exact expressions for the quantum Fisher information for one- and two- mode mixed Gaussian probe states undergoing arbitrary Bogoliubov transformations and illustrated its use on the estimation of the proper acceleration.

By employing a series expansion in the Bogoliubov coefficients around the point $\epsilon_{0}\,=\,0$, we were able to evaluate the quantum Fisher information for the case of one- and two-mode Gaussian probe states. We obtained exact expressions for the quantum Fisher information at point $\epsilon\,=\,0$, and perturbative expressions for $\epsilon\neq 0$. In the single mode case, for a finite temperature, the quantum Fisher information is identically zero at $\epsilon\,=\,0$. This implies that for states which are at some temperature other than absolute zero, one cannot distinguish between two states in the neighbourhood of $\epsilon\,=\,0$. For larger values of $\epsilon$, the quantum Fisher information is non-zero and the quantum Cram\'{e}r-Rao bound is finite. On the other hand, in the case of a thermal two-mode state there is always the possibility of distinguishing between infinitesimally close states in the neighbourhood of $\epsilon\,=\,0$.


Higher squeezing and a high temperature difference of modes significantly improves the precision in estimation. Squeezed states are generally more sensitive to rotations and mode-mixing as well as particle creation when the squeezed state is appropriately oriented. The difference in temperature also helps because any mode-mixing channel given by a non-trivial passive coefficient $\alpha$ will, in general, mix temperatures of different modes. This effect vanishes when modes have the same temperature. Ultimately, this is due to the fact that squeezed thermal states have high variance in energy which, according to the general equation for the quantum Fisher information~\cite{Paris2009a}, leads to a greater precision in estimation.


Our results will enable researchers to evaluate how well space-time parameters, such as the amplitude of gravitational waves, accelerations and local gravitational fields, can be estimated in the presence of background temperature~\cite{Sabin2014a,Sabin2015a}. We observe that strategies involving one- and two-mode probe states exhibit the same exponential gain for large squeezing. However, single mode thermal states do not perform well in the scenario when the channel is mostly mode-entangling, which is a common case in the literature~\cite{Bruschi2012a,Bruschi2013b,Friis2013a,Sabin2014a}.

Our results lead naturally to other important questions. The quantum Fisher information is the optimisation of the classical Fisher information over all possible measurements. Therefore, \emph{what is the optimum measurement for our scheme?} An analysis of the symmetric logarithmic derivative would certainly shed light on this and general knowledge in this direction has already been developed~\cite{Monras2013a}. Furthermore, if the optimal measurement is found to be impractical then an analysis of more realistic measurements, such has homodyne and heterodyne for Gaussian states, could prove fruitful. These questions are left for future work.

\acknowledgments

We would like to thank Alex Monras and Carlos Sab\'{i}n for useful comments and correspondence. I. F. acknowledges support from EPSRC (CAF Grant No. EP/G00496X/2 to I. F.). D. E. B. was partially supported by the I-CORE Program of the Planning and Budgeting Committee and the Israel Science Foundation (grant No. 1937/12), as well as by the Israel Science Foundation personal grant No. 24/12.

\appendix

\section{\label{app1}Real and complex form of first and second moments}

Here we present the connection between the ``real" form of the covariance matrix formalism (commonly used in the literature), and the ``complex" form, used in this paper. Our choice was based on mathematical simplicity.

The real form of the covariance matrix formalism is usually defined with respect to the collection of quadrature operators $\boldsymbol{\hat{Q}}\,\dot{=}\,\boldsymbol{\hat{x}}\oplus\boldsymbol{\hat{p}}\,=\,\{\hat{x}_{1},\hat{x}_{2},\ldots,\hat{p}_{1},\hat{p}_{2},\ldots\}$. The canonical commutation relations of these operators can be conveniently expressed as
\begin{equation}
[\hat{Q}_{m},\hat{Q}_{n}]\,=+i\,\Omega_{mn}\,\mathrm{id}\quad\Rightarrow\quad\boldsymbol{\Omega}\,=\,
\begin{bmatrix}
\boldsymbol{0} & \boldsymbol{I} \\
\boldsymbol{-I} & \boldsymbol{0}
\end{bmatrix}.
\end{equation}
Notice some properties of $\boldsymbol{\Omega}$ are $-\boldsymbol{\Omega}^{2}\,=\,+\boldsymbol{I}$ and $\boldsymbol{\Omega}^{\mathrm{tp}}\,=\,-\boldsymbol{\Omega}$, in contrast to the complex form version $\boldsymbol{K}$. In the real form, the definitions of the first and second moments are,
\begin{subequations}
\begin{align}
\boldsymbol{d}_{R}&\,=\,\mathrm{tr}\,[\hat{\rho}\,\boldsymbol{\hat{Q}}]\,=\,\begin{bmatrix}\boldsymbol{x} \\ \boldsymbol{p}\end{bmatrix},\\
\boldsymbol{\sigma}_{R}&\,=\,\mathrm{tr}\,[\hat{\rho}\,\{\boldsymbol{\hat{Q}},\boldsymbol{\hat{Q}}\}]\,=\,\begin{bmatrix}\boldsymbol{A}_{R} & \boldsymbol{B}_{R} \\ \boldsymbol{B}^{\mathrm{tp}}_{R} & \boldsymbol{C}_{R} \end{bmatrix}.
\end{align}
\end{subequations}
Note that the sub-matrices $\boldsymbol{A}_{R}$ and $\boldsymbol{C}_{R}$ are \emph{not} the reduced states of a subsystem. We can now switch between the two different representations via the transformation,
\begin{equation}
\boldsymbol{\hat{A}}\,=\,\boldsymbol{L}\boldsymbol{\hat{Q}},\quad\quad\boldsymbol{L}\,\dot{=}\,\frac{1}{\sqrt{2}}\,\begin{bmatrix}\boldsymbol{I} & +i\boldsymbol{I} \\ \boldsymbol{I} & -i\boldsymbol{I}\end{bmatrix}.
\end{equation}
The transformation between the two forms is then given by,
\begin{equation}
\label{eqn:appendix_real_to_complex_transformation}
\boldsymbol{d}\,=\,\boldsymbol{L}\boldsymbol{d}_{R},\quad\quad\boldsymbol{\sigma}\,=\,\boldsymbol{L}\boldsymbol{\sigma}_{R}\boldsymbol{L}^{\dag}.
\end{equation}
In terms of the components of the first and second moments, we have the identification,
\begin{equation}
\begin{gathered}
\boldsymbol{\alpha}\,=\,\frac{\boldsymbol{x}+i\,\boldsymbol{y}}{\sqrt{2}},\quad\boldsymbol{X}\,=\frac{\boldsymbol{A}_{R}+\boldsymbol{C}_{R}-i\,\big(\boldsymbol{B}_{R}-\boldsymbol{B}^{\mathrm{tp}}_{R}\big)}{2},\\
\boldsymbol{Y}\,=\,\frac{\boldsymbol{A}_{R}-\boldsymbol{C}_{R}+i\,\big(\boldsymbol{B}_{R}+\boldsymbol{B}^{\mathrm{tp}}_{R}\big)}{2}.
\end{gathered}
\end{equation}
Notice the required conditions $\boldsymbol{X}^{\dag}\,=\,\boldsymbol{X}$ and $\boldsymbol{Y}^{\mathrm{tp}}\,=\,\boldsymbol{Y}$. Collecting these expressions back into the first and second moments gives us the representation,
\begin{equation}
\boldsymbol{d}\,=\,
\begin{bmatrix}
\boldsymbol{d}_{\boldsymbol{a}} \\ \overline{\boldsymbol{d}}_{\boldsymbol{a}}
\end{bmatrix},\quad
\boldsymbol{\sigma}\,=\,\begin{bmatrix}
\boldsymbol{X} & \boldsymbol{Y} \\
\overline{\boldsymbol{Y}} & \overline{\boldsymbol{X}}
\end{bmatrix}.
\end{equation}
This is the source of the notation used in the main body of the paper.

\section{\label{app2}Real and complex form of Bogoliubov coefficients}

Bogoliubov transformations can be represented as a matrix acting on the space of classical Klein-Gordon field solution space or as a matrix acting on the space of canonical field creation and annihilation operators. Algebraically these are, respectively,
\begin{equation}
\boldsymbol{S}\,=\,
\begin{bmatrix}
\boldsymbol{\alpha} & \boldsymbol{\beta} \\
\overline{\boldsymbol{\beta}} & \overline{\boldsymbol{\alpha}}
\end{bmatrix},\quad
\boldsymbol{\tilde{S}}\,=\, \begin{bmatrix}
\overline{\boldsymbol{\alpha}} & -\overline{\boldsymbol{\beta}} \\
-\boldsymbol{\beta} & \boldsymbol{\alpha}
\end{bmatrix}.
\end{equation}
Using the transformation rule in appendix~\ref{eqn:appendix_real_to_complex_transformation}, we can find the real form of Bogoliubov transformation via $\boldsymbol{S}_{R}\,=\,\boldsymbol{L}^{\dag}\boldsymbol{S}\boldsymbol{L}$. The results are
\begin{subequations}
\begin{align}
\boldsymbol{S}_{R}&\,=\,
\begin{bmatrix}
\mathrm{Re}\,\big[\boldsymbol{\alpha}+\boldsymbol{\beta}\big] & -\mathrm{Im}\,\big[\boldsymbol{\alpha}-\boldsymbol{\beta}\big] \\
\mathrm{Im}\,\big[\boldsymbol{\alpha}+\boldsymbol{\beta}\big] & \mathrm{Re}\,\big[\boldsymbol{\alpha}-\boldsymbol{\beta}\big]
\end{bmatrix}, \\
\boldsymbol{\tilde{S}}_{R}&\,=\,
\begin{bmatrix}
\mathrm{Re}\,\big[\boldsymbol{\alpha}-\boldsymbol{\beta}\big] & -\mathrm{Re}\,\big[\boldsymbol{\alpha}+\boldsymbol{\beta}\big] \\
-\mathrm{Im}\,\big[\boldsymbol{\alpha}+\boldsymbol{\beta}\big] & -\mathrm{Im}\,\big[\boldsymbol{\alpha}-\boldsymbol{\beta}\big]
\end{bmatrix}.
\end{align}
\end{subequations}
The matrices $\boldsymbol{S}_{R}$ and $\tilde{\boldsymbol{S}}_{R}$ have been used to define the Bogoliubov transformations in previous literature, such as~\cite{Friis2013a,Sabin2014a,Ahmadi2014a}.

\section{\label{app3}One- and two-mode state elements}

We can define the initial state of the whole system via,
\begin{equation}
\boldsymbol{\tilde{\sigma}}_{0}\,=\,
\begin{bmatrix}
\boldsymbol{X}_{0} && \boldsymbol{Y}_{0} \\
\overline{\boldsymbol{Y}}_{0} && \overline{\boldsymbol{X}}_{0}
\end{bmatrix}
\end{equation}
Note this takes into account any modes of interest and also the remaining ``environment" modes of the system. The general components for the matrices $\boldsymbol{X}$ and $\boldsymbol{Y}$ which constitute a covariance matrix after a general Bogoliubov transformation, given by the quantum channel Eq.~\eqref{eqn:QFI_quantum_channel}, can be written as
\begin{widetext}
\begin{subequations}
\begin{align}
X_{ij}&\,=\,\sum_{a,b}\Big(\alpha_{ia}\,X_{0,ab}\,\overline{\alpha}_{jb}+\beta_{ia}\,\overline{Y}_{0,ab}\overline{\alpha}_{jb}+\alpha_{ia}Y_{0,ab}\overline{\beta}_{jb}+\beta_{ia}\overline{X}_{0,ab}\overline{\beta}_{jb}\Big),\\
Y_{ij}&\,=\,\sum_{a,b}\Big(\beta_{ia}\,\overline{X}_{0,ab}\,\alpha_{jb}+\alpha_{ia}\,Y_{0,ab}\,\alpha_{jb}+\beta_{ia}\,\overline{Y}_{0,ab}\,\beta_{jb}+\alpha_{ia}\,X_{0,ab}\,\beta_{jb}\Big).
\end{align}
\end{subequations}
\end{widetext}
These expressions, coupled with the exact definitions of the quantum Fisher information and Bogoliubov co-efficients, can be used to compute the quantum Fisher information for any state within our quantum field theory framework. The resulting expressions are rather unwieldy and hence we have not written them out explicitly.

\subsection{One-mode covariance matrix elements}

For the two-mode state considered in Eq.~\eqref{eqn:Single_mode_state}, the exact two-mode covariance matrix elements after a general Bogoliubov transformation are given by,
\begin{widetext}
\begin{subequations}
\label{eqn:one_mode_XYsummed}
\begin{align}
X_{mm}&=\nu_{m}\,\Big(\cosh(2r)\,\big(|\alpha_{mm}|^{2}+|\beta_{mm}|^{2}\big)-2\,\mathrm{Re}\big[\alpha_{mm}\overline{\beta}_{mm}\big]\sinh(2r)\Big)+\sum_{a\ne m}\nu_{a}\,\big(|\alpha_{ma}|^{2}+|\beta_{ma}|^{2}\big),\\
Y_{mm}&=\nu_{m}\Big(-2\,\cosh(2r)\,\overline{\alpha}_{mm}\overline{\beta}_{mm}+\big(\overline{\alpha}_{mm}^{2}+\overline{\beta}_{mm}^{2}\big)\sinh(2r)\Big)-2\,\sum_{a\ne m}\nu_{a}\overline{\alpha}_{ma}\overline{\beta}_{ma}.
\end{align}
\end{subequations}
\end{widetext}

\subsection{Two-mode covariance matrix elements}

For the single mode state considered in Eq.~\eqref{eqn:Two_mode_state}, the exact one-mode covariance matrix elements after a general Bogoliubov transformation are given by,
\begin{widetext}
\begin{subequations}
\label{eqn:two_mode_XYsummed}
\begin{align}
\begin{split}X_{ij}&=D_{mn}(\alpha_{im}\overline{\alpha}_{jm}+\beta_{im}\overline{\beta}_{jm})+C_{mn}(\beta_{im}\overline{\alpha}_{jn}+\alpha_{im}\overline{\beta}_{jn})
+D_{nm}(\alpha_{in}\overline{\alpha}_{jn}+\beta_{in}\overline{\beta}_{jn})\\
&\quad\quad+C_{nm}(\beta_{in}\overline{\alpha}_{jm}+\alpha_{in}\overline{\beta}_{jm})
+\sum_{a\neq m,n}\nu_{a}(\alpha_{ia}\overline{\alpha}_{ja}+\beta_{ia}\overline{\beta}_{ja}),
\end{split}\\
\begin{split}Y_{ij}&=D_{mn}(\beta_{im}{\alpha}_{jm}+\alpha_{im}{\beta}_{jm})+C_{mn}(\alpha_{im}{\alpha}_{jn}+\beta_{im}{\beta}_{jn})
+D_{nm}(\beta_{in}{\alpha}_{jn}+\alpha_{in}{\beta}_{jn})\\
&\quad\quad+C_{nm}(\alpha_{in}{\alpha}_{jm}+\beta_{in}{\beta}_{jm})+\sum_{a\neq m,n}\nu_{a}(\beta_{ia}{\alpha}_{ja}+\alpha_{ia}{\beta}_{ja}).
\end{split}
\end{align}
\end{subequations}
\end{widetext}

\bibliographystyle{apsrev}
\bibliography{Metrology_Methods_In_QFT_BiBTeX_ARL}

\end{document}